\documentclass[aps,pra,superscriptaddress,twocolumn,showkeys]{revtex4-2}

\usepackage{graphicx}
\usepackage{amsmath}
\usepackage{amssymb, physics}
\usepackage{bm} 
\usepackage{hyperref} 
\usepackage[T1]{fontenc}
\usepackage[utf8]{inputenc}

\hypersetup{colorlinks=true, linkcolor=blue, citecolor=red, urlcolor=blue}

\begin{document}

\title{From Local Nonclassicality to Entanglement: A Convexity Law for Single-Excitation Dynamics}

\author{Atta ur Rahman}
\affiliation{School of Physics, University of Chinese Academy of Sciences, Yuquan Road 19A, Beijing, 100049, China}
\author{Ao-xiang Liu}
\affiliation{School of Physics, University of Chinese Academy of Sciences, Yuquan Road 19A, Beijing, 100049, China}
\author{M. Y. Abd-Rabbou}
\affiliation{School of Physics, University of Chinese Academy of Sciences, Yuquan Road 19A, Beijing, 100049, China}
\affiliation{Mathematics Department, Faculty of Science, Al-Azhar University, Nassr City 11884, Cairo, Egypt}
\author{Cong-Feng Qiao}
\thanks{\textcolor{blue}{Corresponding author: qiaocf@ucas.ac.cn}}
\affiliation{School of Physics, University of Chinese Academy of Sciences, Yuquan Road 19A, Beijing, 100049, China}

\affiliation{International Centre for Theoretical Physics Asia-Pacific, UCAS, Beijing 100190, China}

\date{\today}

\begin{abstract}
We prove a simple dynamical law for excitation-preserving interactions: the {sum of local Wigner negativities} is upper-bounded by a fixed budget set by the initially excited state. For the single-excitation sector of the XY model (and its beam-splitter analogue), this convexity bound equals the negativity of the seed state and is saturated only when the excitation is fully localized. At intermediate times the sum lies strictly below the bound due to phase-space overlap in local mixtures, quantitatively accounting for entanglement growth as a redistribution of a finite, budgeted resource into shared correlations. We establish the result analytically for two bodies and corroborate it numerically in engineered state-transfer chains, where it reveals a coherence-enabled dark transport: the resource becomes locally invisible while being stored in multi-body coherences. The predicted trajectory of the summed local negativity provides a practical hardware metric: deviations from the ideal, budgeted curve diagnose decoherence and control error.
\end{abstract}
\maketitle

\section{Introduction}
\label{sec:intro}

The power of quantum technologies stems from their ability to process distinct forms of quantum resources, from the local non-classicality of a single system to the shared non-local correlations of entanglement. While the static properties of these resources are increasingly well understood within comprehensive theoretical frameworks~\cite{Chitambar2019, Gour2015}, the fundamental rules governing how one resource is dynamically {interconverted} into another during a physical evolution remain an active and crucial area of investigation~\cite{Dowling2003, Brandao2015}. At the heart of this promise lies the idea of quantum resources, specific properties such as superposition, coherence, magic, and entanglement that fuel applications from secure communication and advanced computation to enhanced sensing~\cite{Nielsen2010, Giovannetti2011, Preskill2018}. Among these, entanglement has long held a special status, with recent theoretical and experimental works continuing to push the boundaries of its generation, certification, and application in multi-qubit systems~\cite{Horodecki2009, Kimble2008, Omran2019, Gong2021}.

Complementing this shared, non-local resource is the non-classicality inherent in {local} states. A prominent phase-space perspective uses the Wigner function~\cite{Wigner1932, Hillery1984}, whose negativity provides an unambiguous and quantifiable signature of non-classicality~\cite{Kenfack2004, Zurek2001}. This has been experimentally demonstrated not only in foundational continuous-variable (CV) systems~\cite{Hofheinz2009, Vlastakis2013} but also more recently in discrete-variable platforms like trapped ions and superconducting circuits~\cite{Fluhmann2019, Campagne-Ibarcq2020}. Far from a theoretical curiosity, Wigner negativity (or, in DV terms, non-stabilizerness/contextuality) is understood to be a critical ingredient for quantum computational advantage within the standard Clifford/stabilizer (DV) and Gaussian (CV) frameworks. According to the Gottesman–Knill theorem, circuits comprising only Clifford gates acting on stabilizer states can be efficiently simulated classically~\cite{Gottesman1998}. Universal quantum computation therefore requires access to non-stabilizer states, or magic states, which are characterized by Wigner negativity~\cite{Bravyi2005, Mari2012, Bravyi2019}. This connection between negativity, contextuality, and computational power establishes Wigner negativity as the {local fuel} for quantum processing, a concept that has been central to recent analyses of fault-tolerance thresholds and the cost of magic state distillation~\cite{Veitch2012, Howard2014, Pashayan2015, Raussendorf2020, Beverland2022}.

While these powerful frameworks excel at characterizing resources as static quantities under restricted operations, a quantitative, dynamical law governing the {interconversion} of one resource into another under continuous-time Hamiltonian evolution has remained elusive. Resource theories define monotones that are non-increasing under free operations, but do not typically describe the precise trajectory of resource exchange under a general Hamiltonian. Conversely, studies of Hamiltonian dynamics often focus on state fidelity or the decay of a resource under noise~\cite{Hashim2021}, rather than a strict accounting of its redistribution. What has been missing is a unifying principle that treats local non-classicality as a {budgeted fuel} for entanglement, quantitatively linking the consumption of the former to the growth of the latter. This is a critical gap, as a predictive, resource-centric understanding of coherent dynamics is essential for designing and benchmarking quantum devices.

{This work closes that gap by establishing a convexity-bounded redistribution principle.} We show that for XY dynamics—the elementary, excitation-exchange interaction realized across leading platforms from superconducting circuits and trapped ions to quantum dots and neutral atoms~\cite{Blais2021, Bruzewicz2019, Loss1998, Saffman2010, Kjaergaard2020}—and for its ideal CV analogue (the beam-splitter), the {sum of local Wigner negativities is constrained by a convexity upper bound}. For ideal unitary evolution in the single-excitation sector, this bound is saturated at endpoints of state transfer. Between them, the sum remains below the bound due to phase-space overlap (cancellation) in local mixtures. Entanglement growth is thereby explained as the coherent apportioning of a finite, budgeted amount of local non-classicality into shared correlations; the {local shares} oscillate out of phase, while their {sum} follows a predictable trajectory dictated by the bound and this overlap effect. We demonstrate this principle analytically in the two-body case and numerically extend it to state-transfer chains, where we uncover a remarkable ``dark transport'' mechanism: the quantum resource becomes locally invisible by encoding itself into multi-body correlations, a process we show is protected by coherence.

A crucial clarification is representation dependence. For odd-prime qudit dimensions, there exist discrete Wigner functions for which stabilizer states and their mixtures are non-negative, implying {zero} local negativity and making our convexity bound trivial in those discrete settings~\cite{Gross2006,Veitch2014}. By contrast, the qubit case $d=2$ is exceptional: one cannot simultaneously have Clifford covariance and universal stabilizer non-negativity~\cite{Gross2006}. To obtain a {non-trivial} and physically transparent budget in our demonstrations, we therefore use a continuous-variable (CV) phase-space embedding, where the single-photon Fock state $\ket{1}$ has strictly positive Wigner negativity. Under the beam-splitter analogue, the {nonzero} sum of local negativities is upper-bounded by this single-photon value. This bound is saturated at the start and end of a swap, with a physically meaningful gap appearing at intermediate times due to phase-space cancellation effects.

We make these scopes explicit and use both settings to illuminate when the bound is informative and how it can be leveraged as a hardware metric. In particular, the {stability of the summed local negativity relative to its ideal trajectory} under target dynamics provides a powerful, resource-centric performance indicator: departures from this ideal behavior directly diagnose nonunitarity due to decoherence and control imperfections. The remainder of the article is organized as follows. In Sec.~\ref{sec:framework} we lay the theoretical groundwork, define the XY/beam-splitter models and resource metrics, and present an analytical derivation of the convexity upper bound. In Sec.~\ref{sec:two_qubit_results} we validate this prediction numerically and provide a phase-space interpretation. Sec.~\ref{sec:n_chain} generalizes to PST chains, and Sec.~\ref{sec:cv_systems} establishes the CV counterpart. Sec.~\ref{sec:conclusion} proposes an experiment and argues that deviations from the ideal trajectory furnish a practical, resource-centric hardware metric.

\section{Theoretical Framework}
\label{sec:framework}

In this section, we establish the theoretical foundation for our analysis. We first introduce the physical model of two interacting qubits and define the key quantum resources—entanglement and non-classicality—that we track. We then present a complete analytical solution for the system's dynamics and, from it, an upper bound governing the redistribution of local non-classicality, providing the resource-budget picture at the heart of this work.

\subsection{The Two-Qubit XY Model}

We consider a closed bipartite quantum system composed of two interacting qubits, which we label A and B. Their coherent dynamics are governed by the isotropic XY interaction Hamiltonian, a fundamental model for spin exchange that is physically realized in various quantum computing architectures~\cite{Blais2021, Bruzewicz2019}. Setting $\hbar=1$, the Hamiltonian is
\begin{equation}
    H = \frac{g}{2}\, (\sigma_x^A \otimes \sigma_x^B + \sigma_y^A \otimes \sigma_y^B),
    \label{eq:hamiltonian}
\end{equation}
which is equivalent to its rotating-wave form
\begin{equation}
    H = g \,(\sigma_+^A \otimes \sigma_-^B + \sigma_-^A \otimes \sigma_+^B).
    \label{eq:hamiltonian_raising_lowering}
\end{equation}
This form makes the physical process transparent: the interaction coherently exchanges a single quantum of excitation. Consequently, the total excitation-number operator,
\begin{equation}
\hat{N}
= \sigma_+^A \sigma_-^A + \sigma_+^B \sigma_-^B
= \tfrac{1}{2}\big(\mathbb{I}-\sigma_z^A\big) + \tfrac{1}{2}\big(\mathbb{I}-\sigma_z^B\big),
\label{eq:number_op_equiv}
\end{equation}
commutes with the Hamiltonian, $[H,\hat{N}]=0$. The dynamics are therefore block-diagonal in excitation number, and a system initialized with a single excitation remains confined to the single-excitation subspace.

We initialize the system at $t=0$ in the separable product state $\ket{\psi(0)}=\ket{0}_A \otimes \ket{1}_B$. In this configuration, a single quantum of excitation—and its associated non-classicality—is localized entirely on qubit B. This initial state defines the total non-classical resource budget, which we denote $\mathcal{N}_1 \equiv \mathcal{N}(\ket{1})$. As we will show, the value of this budget is representation-dependent: for odd-prime qudit dimensions one can choose a discrete Wigner representation with stabilizer non-negativity (yielding $\mathcal{N}_1=0$), whereas in our continuous-variable embedding $\mathcal{N}_1>0$; the qubit case $d=2$ is exceptional and does not admit universal stabilizer non-negativity~\cite{Gross2006}.

\subsection{Measures of Quantum Resources}

To track the interconversion of quantum resources, we employ well-established metrics for non-local correlations (entanglement) and local quantum character (non-classicality). A crucial mathematical property underpinning our central result is the convexity of the Wigner negativity measure.

\paragraph{Entanglement: Concurrence}
For a two-qubit system, entanglement is quantified by the concurrence, $C(t)$~\cite{Wootters1998}. For a pure bipartite state $\ket{\psi(t)}$, it is given by
\begin{equation}
    C(t) \equiv \sqrt{\,2\!\left(1 - \Tr[\rho_A(t)^2]\right)} \,,
    \label{eq:concurrence}
\end{equation}
where $\rho_A(t) = \Tr_B\!\left[\,\ket{\psi(t)}\bra{\psi(t)}\,\right]$ is the reduced density operator of qubit A. Concurrence provides a continuous measure of the non-local quantum correlations shared between the qubits, ranging from $C=0$ (separable) to $C=1$ (maximally entangled).

\paragraph{Non-Classicality: Wigner Negativity}
The non-classical character of the local state of each qubit, $\rho_i$ ($i \in \{A,B\}$), is captured by its Wigner negativity. The Wigner function, $\mathcal{W}(\alpha)$, is a quasiprobability distribution over a phase space that provides a complete state description~\cite{Wigner1932, Hillery1984}. The presence of negative values in this distribution is an unambiguous signature of non-classicality~\cite{Kenfack2004}. The Wigner negativity, $\mathcal{N}(\rho_i)$, quantifies this non-classical volume:
\begin{equation}
    \mathcal{N}(\rho_i) = \frac{1}{2} \left( \int |\mathcal{W}_i(\alpha)| \, d^2\alpha - 1 \right).
    \label{eq:negativity}
\end{equation}
We use the convention $\int \mathcal{W}_i(\alpha)\,d^2\alpha=1$, so classical states have $\mathcal{N}(\rho_i)=0$. (For discrete phase spaces, the integral is understood as a normalized sum.) Wigner negativity is a necessary resource for quantum computational speedups~\cite{Mari2012, Howard2014}. Crucially for our analysis, negativity is a {convex functional}: for any mixture $\rho = \sum_k p_k \rho_k$, it obeys the inequality $\mathcal{N}(\rho) \le \sum_k p_k \mathcal{N}(\rho_k)$. This property is the mathematical foundation of our resource bound.

\paragraph{Stabilizer Non-Classicality: Mana}
While Wigner negativity provides a phase-space picture of non-classicality, related stabilizer-based measures are also informative. The {sum negativity} (the $L^1$ excess of the discrete Wigner function) is convex; by contrast, its logarithmic variant, mana, is a useful monotone but {not convex}. Our convexity-based bound {does not} extend to mana; replacing sum negativity by mana generally invalidates the inequality~\cite{Veitch2014}. We therefore do not claim an analogous convexity law for mana.

\subsection{Isomorphism to the Beam-Splitter Model and Choice of Wigner Representation}
\label{sec:cv_isomorphism}

A key aspect of our analysis is the choice of Wigner representation, which determines whether the negativity budget $\mathcal{N}_1$ is non-trivial. The single-excitation dynamics of the XY Hamiltonian are isomorphic to the single-photon dynamics of two bosonic modes coupled by a beam-splitter Hamiltonian, $H_{\text{BS}} = g(a^\dagger b + a b^\dagger)$. The qubit states $\ket{0}$ and $\ket{1}$ map directly to the Fock states $\ket{0}$ and $\ket{1}$ of the bosonic modes.

This isomorphism allows us to leverage the standard continuous-variable Wigner function, which provides a physically grounded and representationally clean framework. In this CV representation, the ground state (vacuum) $\ket{0}$ is a Gaussian state with a non-negative Wigner function, hence $\mathcal{N}(\ket{0})=0$. The excited state (single photon) $\ket{1}$ is a non-Gaussian state with a Wigner function exhibiting significant negative volume, yielding a strictly positive budget, $\mathcal{N}_1 = \mathcal{N}(\ket{1}) > 0$. While discrete Wigner functions exist for qubits where $\mathcal{N}(\ket{1})\neq0$, they may lack desirable properties like full Clifford covariance. Our choice of the isomorphic CV model—termed a "CV embedding"—thus enables a non-trivial demonstration of the convexity bound with a clear physical interpretation. Unless stated otherwise, all numerical demonstrations and plots in this work employ this CV Wigner function.

\subsection{Analytical Dynamics and the Convexity-Bounded Redistribution Law}
\label{sec:analytical}

The exact solvability of the XY Hamiltonian within the single-excitation sector allows us to derive a precise analytical expression for the system's dynamics and, from it, a tight upper bound governing the redistribution of local non-classicality.

\paragraph{Derivation of the Time-Evolved State}
As established, for the initial state $\ket{\psi(0)} = \ket{01}$, the evolution is restricted to the subspace spanned by $\{\ket{01}, \ket{10}\}$. Within this subspace, the Hamiltonian is $H_\text{sub} = g\,\sigma_x$, and the time-evolution operator is $U_\text{sub}(t) = \cos(gt)\,\mathbb{I} - i \sin(gt)\,\sigma_x$. Applying this to the initial state yields the time-evolved global state:
\begin{equation}
    \ket{\psi(t)} = \cos(gt)\ket{01} - i\sin(gt)\ket{10}.
    \label{eq:analytical_state}
\end{equation}
This state describes a coherent oscillation of the excitation between the two qubits, with a full swap period of $T = \pi/g$ (implying maximal entanglement at $t=T/4$). The corresponding concurrence evolves as
\begin{equation}
    C(t) = \big|\sin(2gt)\big|. \label{eq:Ct_closed}
\end{equation}

\paragraph{Convexity Upper Bound for Local Non-Classicality}
From the global state, we derive the local reduced density matrices by tracing out the other subsystem:
\begin{align}
    \rho_A(t) &= \cos^2(gt) \ket{0}\bra{0} + \sin^2(gt) \ket{1}\bra{1}, \label{eq:rho_A}\\
    \rho_B(t) &= \sin^2(gt) \ket{0}\bra{0} + \cos^2(gt) \ket{1}\bra{1}. \label{eq:rho_B}
\end{align}
While the global state is pure, each local subsystem evolves as a classical mixture of the ground state $\ket{0}$ and the excited state $\ket{1}$. This structure is key. We now apply the convexity property of Wigner negativity. Since the ground state $\ket{0}$ is classical (a stabilizer state or vacuum state), its Wigner function is non-negative, and thus $\mathcal{N}(\ket{0})=0$. Applying the convexity rule to the local states gives the individual bounds:
\begin{align}
    \mathcal{N}_A(t) &\le \sin^2(gt)\, \mathcal{N}(\ket{1}) + \cos^2(gt)\,\mathcal{N}(\ket{0}) = \sin^2(gt)\, \mathcal{N}_1, \label{eq:NA_bound}\\
    \mathcal{N}_B(t) &\le \cos^2(gt)\, \mathcal{N}(\ket{1}) + \sin^2(gt)\,\mathcal{N}(\ket{0}) = \cos^2(gt)\, \mathcal{N}_1. \label{eq:NB_bound}
\end{align}
Summing these individual constraints, we arrive at our central theoretical result: a tight convexity upper bound on the total available local negativity.
\begin{equation}
    \sum_{i=A,B} \mathcal{N}_i(t)
    \;\le\; \big(\sin^2(gt)+\cos^2(gt)\big)\,\mathcal{N}_1
    \;=\; \mathcal{N}_1.
    \label{eq:convexity_bound_law}
\end{equation}
This inequality proves that the sum of local non-classicalities at any time cannot exceed the initial resource budget. The physical process of entanglement generation is thus framed as the redistribution of a resource under a firm budget constraint.

\noindent\textit{Representation Note.} For odd-prime qudit dimensions, one can choose a discrete Wigner representation in which stabilizer states (and their mixtures) are non-negative~\cite{Gross2006,Veitch2014}; in that case the budget $\mathcal{N}_1$ vanishes and Eq.~\eqref{eq:convexity_bound_law} is trivial. The qubit case $d=2$ is exceptional and does not admit a Clifford-covariant Wigner representation with universal stabilizer non-negativity~\cite{Gross2006}. To make the bound non-trivial and experimentally relevant, we analyze the isomorphic single-excitation manifold of two bosonic modes under a beam-splitter (our “CV embedding”), where $\ket{1}$ has strictly positive negativity, $\mathcal{N}_1>0$.

This representation dependence is physically informative. Our primary use of a CV embedding, where the single-photon budget $\mathcal{N}_1$ is strictly positive, provides a non-trivial and experimentally relevant framework for analyzing resource conversion in leading platforms like superconducting circuits and quantum optics, where Fock states are a canonical non-Gaussian resource. The fact that the bound becomes trivial in certain discrete formalisms is itself an insight: it reveals that those specific mathematical constructions are blind to the non-stabilizerness that fuels these dynamics, underscoring the importance of choosing a physically appropriate phase-space representation.
\section{Dynamics and Convexity-Bounded Behavior in the Two-Qubit System}
\label{sec:two_qubit_results}

\begin{figure*}[ht]
 \centering
 \includegraphics[width=\linewidth]{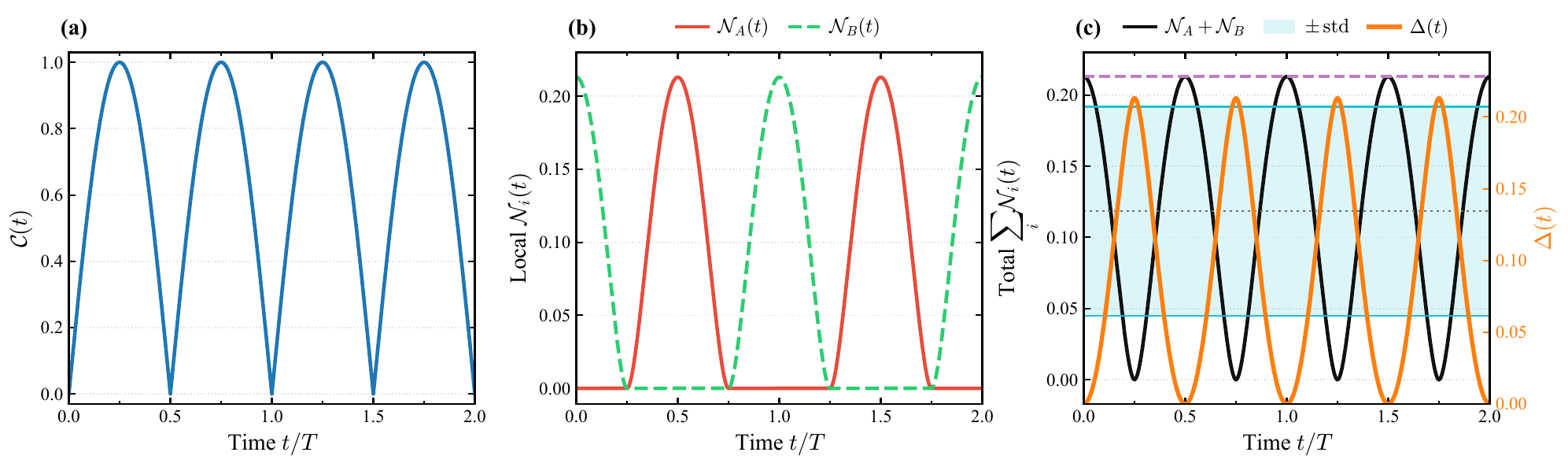}
 \caption{Numerical simulation of resource dynamics and the convexity bound in the two-subsystem exchange model (CV embedding), initialized in $\ket{\psi(0)}=\ket{0}_A\ket{1}_B$. (a) The concurrence $\,\mathcal{C}(t)\,$ oscillates between 0 and 1. (b) Local Wigner negativities $\mathcal{N}_A(t)$ and $\mathcal{N}_B(t)$ are out of phase, evidencing redistribution. (c) The summed local negativity $\mathcal{N}_{\mathrm{tot}}(t)=\mathcal{N}_A(t)+\mathcal{N}_B(t)$ (left axis) remains below the convexity bound $\mathcal{N}_1$ (dashed), saturating it only when the excitation is fully localized on either subsystem (e.g., at times $t/T\in\{0, 0.5, 1.0, \dots\}$). The shaded band shows $\pm1\sigma$ around the mean of $\mathcal{N}_{\mathrm{tot}}(t)$. The right axis plots the tracking gap $\Delta(t)=\max\!\big[0,\mathcal{N}_1-\mathcal{N}_{\mathrm{tot}}(t)\big]$, which quantifies the reduction due to overlap cancellation at intermediate times.}
 \label{fig:dynamics_corrected}
\end{figure*}

We now validate the analytical predictions from Sec.~\ref{sec:analytical} by numerically simulating the system's coherent quantum dynamics. The time-dependent Schrödinger equation for the XY Hamiltonian [Eq.~\eqref{eq:hamiltonian}] was solved using high-precision numerical integrators available in the QuTiP framework~\cite{Johansson2012}, with the system initialized in the state $\ket{\psi(0)} = \ket{0}_A \ket{1}_B$. The results, presented in Fig.~\ref{fig:dynamics_corrected}, confirm that the {sum of local negativities follows a predictable trajectory below the convexity upper bound}.

\noindent{Scope of demonstrations.} As established in Sec.~\ref{sec:cv_isomorphism}, unless explicitly stated otherwise, all non-trivial negativity computations and plots use the continuous-variable (CV) Wigner function corresponding to the isomorphic beam-splitter model. We refer to this as the "CV embedding".

\paragraph{Numerical Method and Tolerances}
We employ QuTiP~\cite{Johansson2012} with adaptive-step ordinary differential equation solvers to ensure high-fidelity propagation of the state vector. For the CV embedding runs shown, we truncate each oscillator's Hilbert space at a dimension of $N_{\text{dim}}=20$. To ensure the robustness of our results, we verify convergence of all observables with respect to this truncation dimension.

\paragraph{Resource Dynamics}
Figure~\ref{fig:dynamics_corrected}(a) shows the concurrence, $C(t)$, which matches the analytical prediction $C(t)=|\sin(2gt)|$ [cf. Eq.~\eqref{eq:Ct_closed}]. This confirms that the XY interaction acts as a perfect entangling channel. The underlying resource dynamics are detailed in Fig.~\ref{fig:dynamics_corrected}(b) (CV embedding), which tracks the local Wigner negativity of subsystem A ($\mathcal{N}_A$) and subsystem B ($\mathcal{N}_B$). The negativities exhibit clear anti-phase oscillations, visually representing a coherent flow of the non-classical resource from the initially excited subsystem B to subsystem A and back again. This anti-correlation is perfectly consistent with a redistribution under the individual convexity bounds of Eqs.~\eqref{eq:NA_bound}–\eqref{eq:NB_bound}.

\paragraph{Convexity-Bounded Sum and the Tracking Gap}
The central behavior is highlighted in Fig.~\ref{fig:dynamics_corrected}(c). This panel plots the total local Wigner negativity,
\begin{equation}
    \mathcal{N}_{\mathrm{tot}}(t) \equiv \sum_{i=A,B} \mathcal{N}_i(t),
    \label{eq:Ntot_def}
\end{equation}
alongside the convexity upper bound from Eq.~\eqref{eq:convexity_bound_law}. In the CV embedding, the numerically computed $\mathcal{N}_{\mathrm{tot}}(t)$ follows a distinct trajectory that is saturated at the swap points ($t/T \in \{0, 0.5, 1, \dots\}$) but dips visibly below the bound at intermediate times, reaching a minimum at the point of maximum entanglement ($t/T \in \{0.25, 0.75, \dots\}$). This "tracking gap" is a genuine physical effect, not a numerical artifact. Between swap points, $\mathcal{N}_A(t)+\mathcal{N}_B(t) < \mathcal{N}_1$. This reduction occurs because the local states are mixtures, e.g., $\rho_A(t) = \cos^2(gt)\ket{0}\bra{0} + \sin^2(gt)\ket{1}\bra{1}$. The positive, Gaussian Wigner function of the $\ket{0}$ component overlaps and cancels the negative lobes of the Wigner function of the $\ket{1}$ component, thereby reducing the total integrated negative volume ($L^1$ norm). This ‘overlap cancellation’ explains the observed tracking gap and is a direct consequence of the phase-space structure of the states involved.

\paragraph{Quantifying the Tracking Gap.}
The tracking gap, $\Delta(t)\equiv \mathcal{N}_1-\mathcal{N}_{\mathrm{tot}}(t)\ge 0$, is a key feature of the dynamics. In our simulations, the gap is visibly non-zero between swap points; its magnitude depends on the specific Wigner representation and phase-space grid but reflects the fundamental effect of overlap cancellation. This ideal trajectory, including the gap, serves as the baseline for our proposed hardware metric.

\subsection{Phase-Space Interpretation of Resource Conversion}
\label{subsec:phase_space}

\begin{figure*}[ht]
 \centering
 \includegraphics[width=0.95\textwidth]{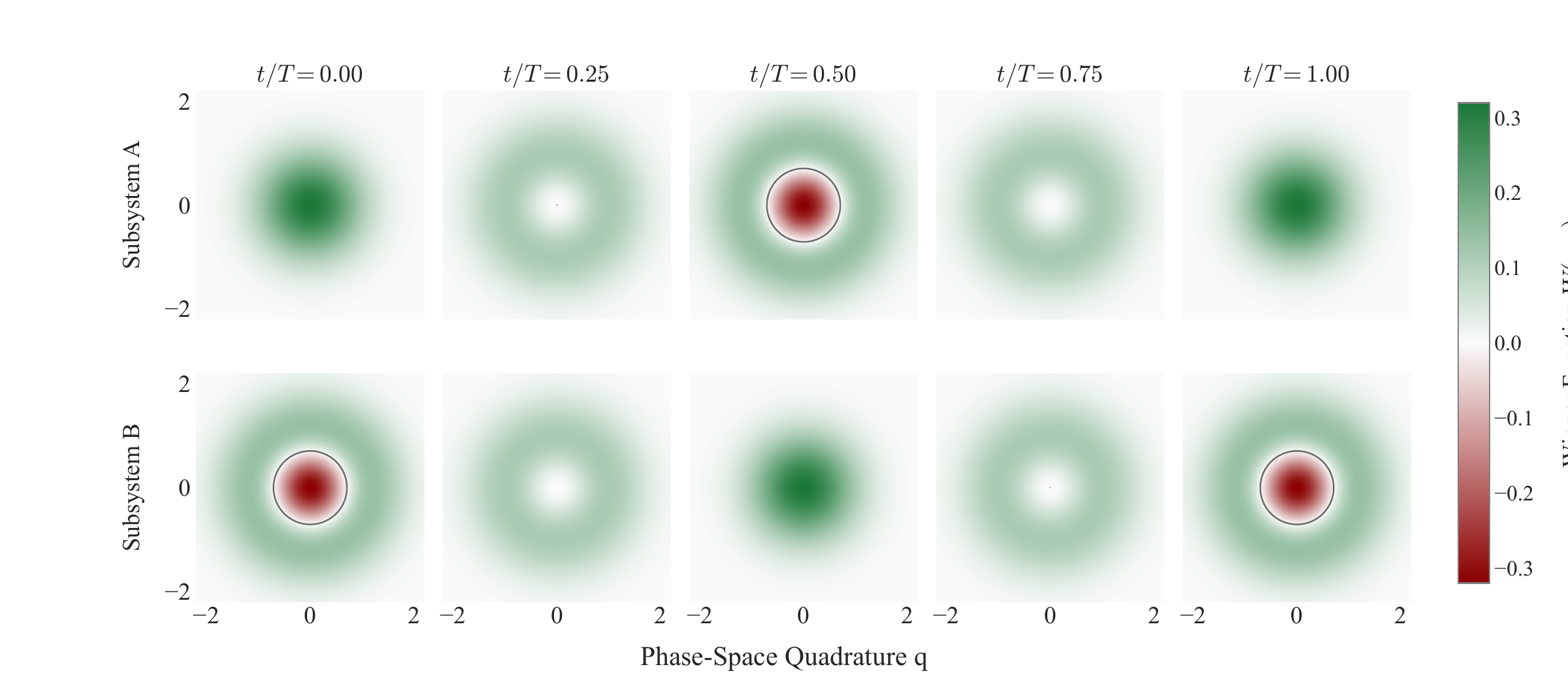}
 \caption{Time-resolved phase-space visualization of resource conversion (CV embedding). Wigner functions for subsystem A (top row) and subsystem B (bottom row) are shown. The sequence illustrates the coherent transfer of non-classicality, initially localized on subsystem B ($t/T=0.00$), through a maximally entangled state where the non-classicality is shared and reduced by overlap cancellation ($t/T=0.25$), to its complete localization on subsystem A ($t/T=0.50$). This demonstrates the redistribution of a {bounded} resource whose sum follows a predictable trajectory below the convexity bound.}
 \label{fig:wigner}
\end{figure*}

To provide a physical interpretation for the convexity-bounded picture, we now analyze the system's dynamics in phase space. Figure~\ref{fig:wigner} presents snapshots of the Wigner functions for subsystems A and B at key moments during the evolution (CV embedding), visually demonstrating how local non-classicality is redistributed under the budget constraint.

\paragraph{Initial State ($t=0$)}
At the initial time, the system is in the separable state $\ket{\psi(0)} = \ket{0}_A \otimes \ket{1}_B$. The local states are pure: $\rho_A(0) = \ket{0}\bra{0}$ and $\rho_B(0) = \ket{1}\bra{1}$. In the CV embedding, subsystem A's Wigner function, $\mathcal{W}_A$, is non-negative (a Gaussian), while subsystem B's Wigner function, $\mathcal{W}_B$, exhibits the negative features of the single-photon Fock state. All non-classicality is initially localized on subsystem B, and the sum of local negativities saturates the upper bound, $\mathcal{N}_{\mathrm{tot}}(0) = \mathcal{N}_1$.

\paragraph{Maximally Entangled State ($t/T=1/4$)}
At a quarter-period ($t = T/4 = \pi/(4g)$), the system reaches a maximally entangled Bell state. The local states become identical statistical mixtures: $\rho_A(T/4) = \rho_B(T/4) = \tfrac{1}{2}(\ket{0}\bra{0} + \ket{1}\bra{1})$. The resulting Wigner functions are identical mixtures, $\mathcal{W}_{A/B} = \tfrac{1}{2}\big(\mathcal{W}_{\ket{0}} + \mathcal{W}_{\ket{1}}\big)$. As seen in Fig.~\ref{fig:wigner}, the positive Gaussian peak of the $\mathcal{W}_{\ket{0}}$ component now overlaps with the central negative region of the $\mathcal{W}_{\ket{1}}$ component. This overlap partially "fills in" the negative volume, reducing the negativity of each local state to less than half of the initial budget. This is the origin of the tracking gap.

\paragraph{Transferred State ($t/T=1/2$)}
By a half-period ($t=T/2$), a full state swap has occurred, and the global state is proportional to $\ket{1}_A \otimes \ket{0}_B$. The non-classical feature—the Wigner negativity—is now entirely localized on subsystem A, and the sum again saturates the bound. The evolution confirms that the XY interaction does not alter the intrinsic phase-space structure of the resource state itself; rather, it coherently modulates the classical probabilities of the local mixture, thereby shuttling a {bounded} budget of Wigner negativity back and forth between the subsystems.

\section{Generalization to N-qubit Chains}
\label{sec:n_chain}

To test whether the convexity-bounded redistribution principle is a general feature of excitation-preserving dynamics, we extend our analysis to an $N$-qubit spin chain. We consider a chain specifically designed for perfect state transfer (PST)~\cite{Christandl2004}, allowing us to test the validity of the bounded redistribution picture beyond simple pairwise interactions.

\subsection{The Perfect State Transfer Model}

We consider a one-dimensional chain of $N$ qubits with nearest-neighbor XY-type interactions, governed by the Hamiltonian
\begin{equation}
    H_{\text{PST}} \;=\; \sum_{k=0}^{N-2} J_k \Big(\sigma_+^{(k)}\sigma_-^{(k+1)} + \sigma_-^{(k)}\sigma_+^{(k+1)}\Big),
    \label{eq:H_PST}
\end{equation}
with engineered couplings $J_k = g\sqrt{(k+1)(N-k-1)}$~\cite{Christandl2004}. This Hamiltonian conserves total excitation number, $[H_{\text{PST}}, \hat{N}] = 0$, confining the dynamics of a single initial excitation. The local reduced state of each qubit therefore remains a classical mixture of $\ket{0}$ and $\ket{1}$, allowing the convexity bound to be generalized to the entire chain.

\subsection{Numerical Results: Coherence-Protected Resource Transport}

\begin{figure*}[htbp]
  \centering
  \includegraphics[width=\textwidth]{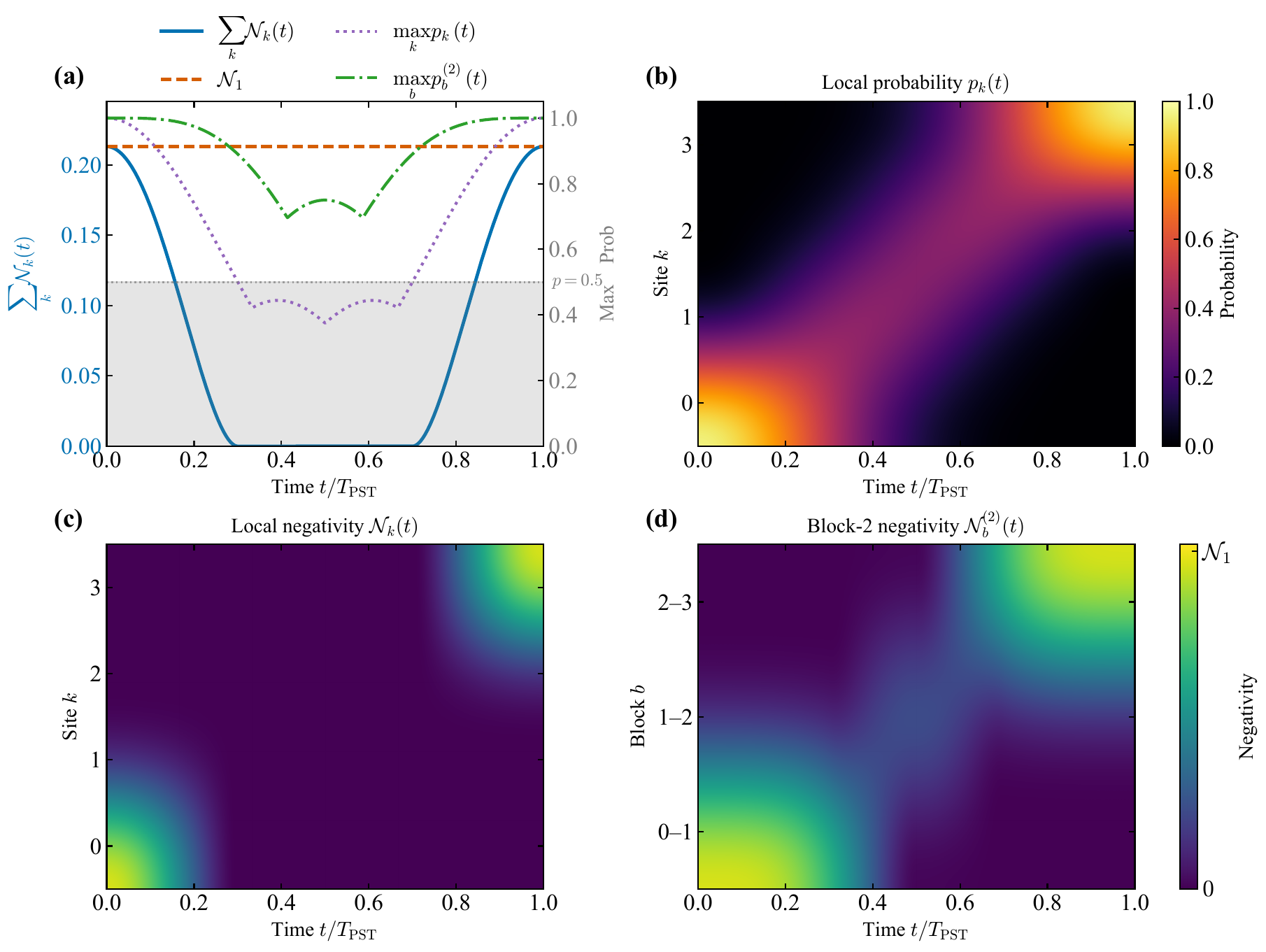}
  \caption{{Convexity-bounded redistribution and coherence-protected transport in a 4-qubit PST chain.}
  {(a)} Resource dynamics summary. The total local negativity ($\sum_k \mathcal{N}_k$, solid blue) remains below the initial resource budget ($\mathcal{N}_1$, orange dashed), vanishing during mid-transport. This disappearance is explained by the maximum single-site probability ($\max_k p_k$, purple dotted) dropping into the negativity-suppressed region ($p < 0.5$, shaded grey). In contrast, the maximum 2-qubit block probability ($\max_b p_b^{(2)}$, green dash-dotted) remains high, sustaining the resource.
  {(b)} Heatmap of the underlying excitation probability ($p_k$) showing perfect state transfer from site 0 to site 3.
  {(c)} Heatmap of local negativity ($\mathcal{N}_k$) revealing a {“dark transport corridor”}: the resource is only visible at the chain ends, appearing to be lost during transit.
  {(d)} Heatmap of block-2 negativity ($\mathcal{N}_b^{(2)}$) {resolves the puzzle}, showing the resource propagating as a continuous, bright wave. The negativity is not lost but is protected in the correlations between qubits, becoming visible only when examining a sufficiently large, coherent block.}
  \label{fig:n_chain_dynamics}
\end{figure*}

We performed a numerical simulation of a chain with $N=4$ qubits, initialized in the state $\ket{\psi(0)} = \ket{1000}$. The results, summarized in Fig.~\ref{fig:n_chain_dynamics}, reveal a rich interplay between resource conservation, localization, and coherence.

\paragraph{Bounded Trajectory in the N-qubit System}
As shown in Fig.~\ref{fig:n_chain_dynamics}(a), the chain-wide sum of local negativities, $\mathcal{N}_{\mathrm{tot}}^{\mathrm{chain}}(t) \equiv \sum_{k=0}^{N-1} \mathcal{N}_k(t)$, adheres strictly to the convexity-bounded redistribution principle. The sum begins at the total resource budget $\mathcal{N}_1$, saturates it again only at the end of the transfer, and remains below it at all intermediate times, consistent with the two-qubit case.

\paragraph{The Puzzle of ``Dark Transport''}
While the total budget is conserved, the local manifestation of the resource is striking. The heatmap of local negativity in Fig.~\ref{fig:n_chain_dynamics}(c) shows that $\mathcal{N}_k$ is non-zero only at the endpoints of the chain. During the transport phase, the chain's interior forms a ``dark transport corridor'' where the resource appears to have vanished completely. The reason is quantified in Fig.~\ref{fig:n_chain_dynamics}(a): the excitation becomes so delocalized (visualized in the probability transfer of Fig.~\ref{fig:n_chain_dynamics}(b)) that the maximum probability on any single site ($\max_k p_k$) falls below the $p=0.5$ threshold required for Wigner negativity. From a purely local perspective, the resource is lost.

\paragraph{Solution: The Resource Hides in Correlations}
The puzzle is resolved by examining a larger subsystem. Figure~\ref{fig:n_chain_dynamics}(d) plots the negativity of contiguous 2-qubit blocks, $\mathcal{N}_b^{(2)}$. This ``block-level'' view reveals the hidden resource, which now propagates as a continuous, bright wave across the chain. The resource was never lost; it transformed from a local, single-body property into a shared, non-local property encoded in the quantum coherence between adjacent qubits. As confirmed in Fig.~\ref{fig:n_chain_dynamics}(a), the probability summed over a 2-qubit block ($\max_b p_b^{(2)}$) remains high enough to sustain negativity. This phenomenon is a direct visualization of {coherence protecting a non-classical resource}, shuttling it losslessly through the many-body system by storing it in correlations.

\section{Analysis in the Native Continuous-Variable Framework}
\label{sec:cv_systems}

We now investigate the convexity-bounded redistribution picture in its most natural setting: the direct continuous-variable (CV) analogue. While the preceding sections have used the CV Wigner function via an isomorphism (the "CV embedding"), this section analyzes the system directly as coupled bosonic modes. This approach confirms the universality of the principle and allows us to highlight features specific to native CV simulations, such as the effect of Hilbert space truncation. Here, the single-photon Fock state provides a concrete, experimentally relevant non-classical {budget}, and the dynamics are governed by the beam-splitter Hamiltonian.

\subsection{The Coupled Oscillator Model}
The CV analogue of the XY interaction is the beam-splitter Hamiltonian,
\begin{equation}
    H_{\text{CV}} = g \,(a^\dagger b + a b^\dagger),
    \label{eq:cv_hamiltonian}
\end{equation}
where $a^\dagger$ and $b^\dagger$ are creation operators for modes A and B. This Hamiltonian conserves the total excitation number, $\hat{N}_{\text{tot}} = a^\dagger a + b^\dagger b$, and produces the same single-excitation dynamics as Eq.~\eqref{eq:analytical_state}. We simulate the initial state $\ket{\psi(0)} = \ket{0}_A \otimes \ket{1}_B$, where $\ket{n}$ denotes a Fock state. The convexity of negativity implies an {upper bound} on the sum of local negativities given by the single-photon budget $\mathcal{N}(\ket{1})$.

\subsection{Numerical Results and Interpretation}

\begin{figure*}[htbp]
 \centering
 \includegraphics[width=\textwidth]{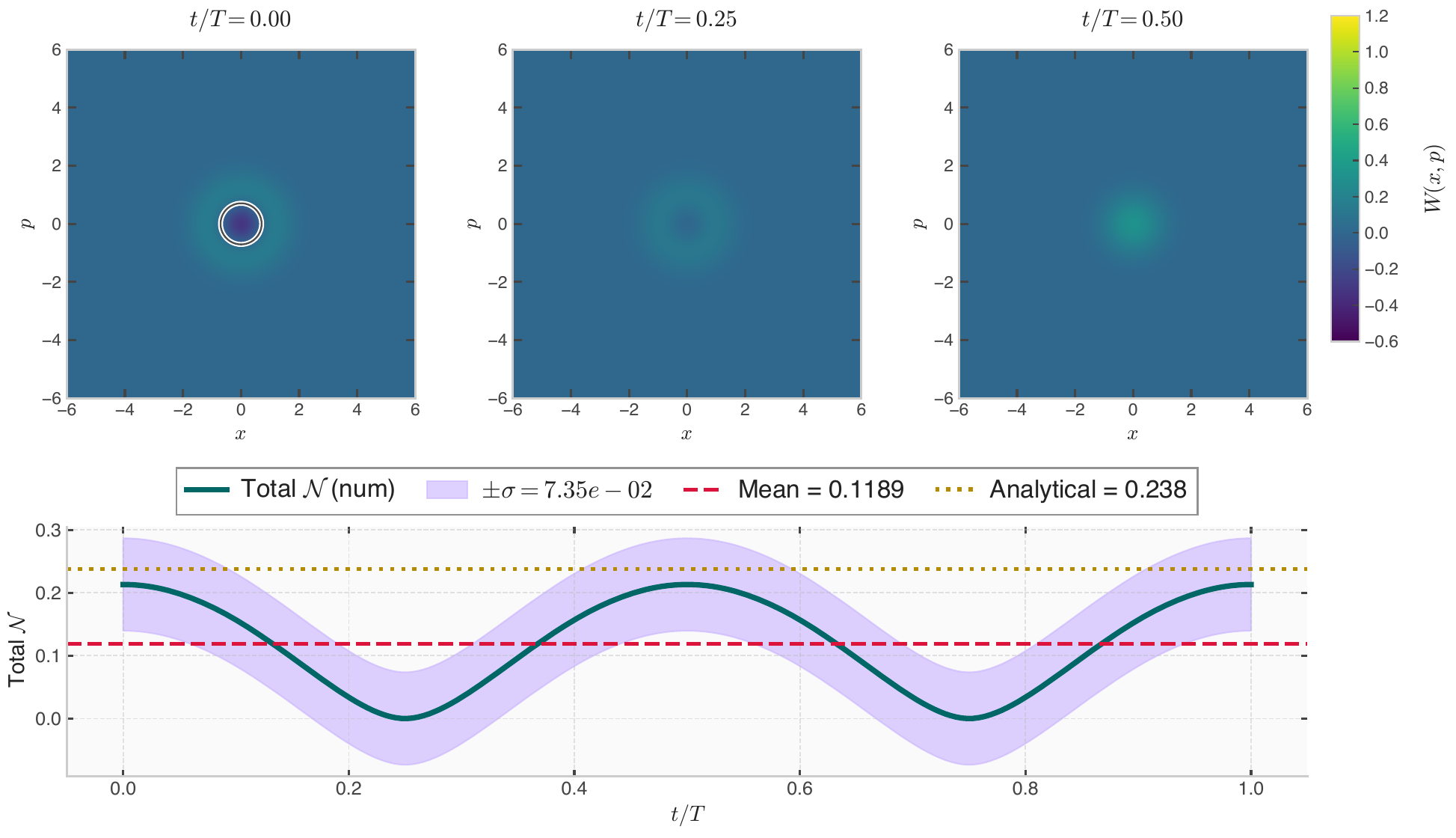}
 \caption{Resource transfer and the convexity bound in a continuous-variable system. Top: Wigner function snapshots of mode B show the transfer of the non-classical Fock state $\ket{1}$ to the vacuum state $\ket{0}$. Bottom: Summed local Wigner negativity $\mathcal{N}^{\mathrm{CV}}_{\mathrm{tot}}(t)$ over one period. The numerically computed trajectory (blue line) lies below the convexity upper bound (gray dotted line), saturating it only at the swap points. The gap at intermediate times is due to physical overlap cancellation, while the small superimposed oscillations are a numerical artifact of Hilbert-space truncation.}
 \label{fig:cv_dynamics}
\end{figure*}

We simulate the system by truncating the Hilbert space of each oscillator at $N_{\text{dim}}=20$. The results are summarized in Fig.~\ref{fig:cv_dynamics}.

\paragraph{Bounded Trajectory and Numerical Artifacts}
The bottom panel of Fig.~\ref{fig:cv_dynamics} plots the total local Wigner negativity, $\mathcal{N}^{\mathrm{CV}}_{\mathrm{tot}}(t) \equiv \mathcal{N}_A(t)+\mathcal{N}_B(t)$. The numerical trajectory clearly shows saturation at the swap points and a tracking gap at intermediate times, perfectly consistent with the physical picture of overlap cancellation. The small, high-frequency oscillations superimposed on this trajectory are a well-understood artifact of Hilbert-space truncation, which subtly perturbs the ideal dynamics. Increasing $N_{\text{dim}}$ systematically reduces the amplitude of these artifactual oscillations, while the overall shape of the trajectory, including the tracking gap, remains. This provides strong numerical validation that the bounded redistribution principle, complete with the physical effect of overlap cancellation, holds true in the ideal CV limit.

\section{Universality of the Redistribution Principle}
\label{sec:universality}

Our analysis thus far has focused on the redistribution of Wigner negativity sourced from an initial single-photon Fock state $\ket{1}$. A crucial question is whether the convexity-bounded redistribution is a specific feature of this state or a more general principle of excitation-preserving dynamics. To answer this, we now investigate the same beam-splitter evolution for different initial non-Gaussian "seed" states, chosen for their distinct phase-space structures and relevance in quantum information.

\subsection{Non-Gaussian Seed States}

In addition to the single-photon state $\ket{1}$, we consider two other canonical non-classical states as the initial resource on mode B:

\paragraph{1. Odd Cat State.} Schrödinger cat states are superpositions of coherent states, forming a cornerstone of CV quantum information and error correction~\cite{Ourjoumtsev2007, Leghtas2015}. We use an odd cat state, defined as $\ket{\psi_{\text{cat}}} \propto (\ket{\alpha} - \ket{-\alpha})$, which is known to possess significant Wigner negativity in a distinct, interference-based pattern in phase space~\cite{Vlastakis2013}.

\paragraph{2. Squeezed Fock State.} Squeezed states of light are a fundamental non-classical resource, particularly for quantum metrology~\cite{Giovannetti2011}. By applying a squeezing operator $S(r)$ to the single-photon state, we create a squeezed Fock state, $S(r)\ket{1}$. This state exhibits a distorted, elliptical Wigner function with negative regions, representing a different class of non-Gaussianity from both Fock and cat states~\cite{Weedbrook2012}.

For each of these initial states, $\ket{\psi(0)} = \ket{0}_A \otimes \ket{\psi_{\text{seed}}}_B$, the dynamics under the beam-splitter Hamiltonian (Eq.~\ref{eq:cv_hamiltonian}) are solved. While the full dynamics are more complex than the two-level oscillation, the local states on each mode, $\rho_A(t)$ and $\rho_B(t)$, are given by mixtures of the vacuum and the time-evolved seed state. Specifically, if $U_{\text{BS}}(t)$ is the beam-splitter unitary, the states on mode A and B are $\rho_A(t) = \text{Tr}_B[U_{\text{BS}}(t)\ket{0\psi_{\text{seed}}}\bra{0\psi_{\text{seed}}}U_{\text{BS}}^\dagger(t)]$ and similarly for $\rho_B(t)$. Due to the linearity of the interaction, these local states are mixtures of the form $\rho_A(t) = \cos^2(gt)\ket{0}\bra{0} + \sin^2(gt)\rho_{\text{seed}}^{\text{eff}}$ and $\rho_B(t) = \sin^2(gt)\ket{0}\bra{0} + \cos^2(gt)\rho_{\text{seed}}^{\text{eff}}$, where $\rho_{\text{seed}}^{\text{eff}}$ is an effective (potentially transformed) seed state. The convexity bound thus continues to hold, with the initial budget $\mathcal{N}_{\text{seed}} = \mathcal{N}(\ket{\psi_{\text{seed}}})$ providing the upper limit.

\subsection{Universal Redistribution Dynamics}

\begin{figure*}[htbp]
  \centering
  \includegraphics[width=\textwidth]{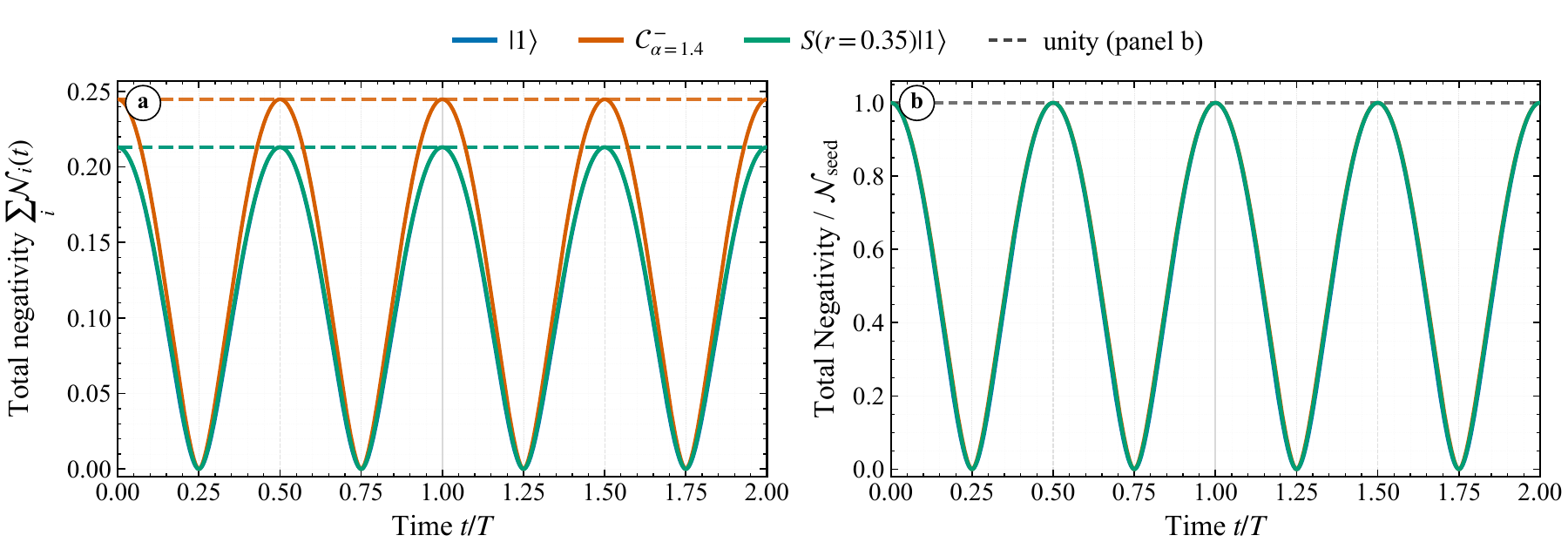}
  \caption{Universality of bounded–budget redistribution for distinct non-Gaussian seed states. 
  (a) Absolute sum of local Wigner negativities, $\sum_i \mathcal{N}_i(t)$, for three seeds prepared on mode $B$: single photon $|1\rangle$ (blue), odd cat $\mathcal{C}^-_{\alpha=1.4}$ (orange), and squeezed $|1\rangle$ with $r=0.35$ (green). Thin dashed lines indicate each seed’s initial budget $\mathcal{N}_{\rm seed}$. Time is shown in units of the exchange period $T$. 
  (b) The same trajectories normalized by their own $\mathcal{N}_{\rm seed}$ collapse onto a common curve, indicating that the redistribution dynamics are largely independent of the seed’s detailed structure.}
  \label{fig:seed_comparison}
\end{figure*}

We numerically compute the trajectory of the summed local negativity, $\sum_i \mathcal{N}_i(t)$, for each of the three seed states. The results are presented in Fig.~\ref{fig:seed_comparison}.

\paragraph{Absolute Dynamics.} The top panel, Fig.~\ref{fig:seed_comparison}(a), shows the absolute summed local negativity. As expected, each seed state starts with a different non-classical budget, $\mathcal{N}_{\text{seed}}$, indicated by the corresponding dashed line. The cat state chosen has the largest budget, followed by the squeezed state, and then the single photon. For all three cases, the dynamics are qualitatively identical: the total local negativity is maximal when the resource is localized on one mode (at $t/T = 0, 0.5, 1, \dots$) and dips significantly at the points of maximal entanglement ($t/T = 0.25, 0.75, \dots$). This confirms that the convexity-bounded redistribution, including the physical effect of a tracking gap due to overlap cancellation, is not unique to the single-photon state.

\paragraph{Normalized Dynamics and a Universal Principle.} The bottom panel, Fig.~\ref{fig:seed_comparison}(b), reveals a deeper, more powerful structure. Here, each trajectory has been normalized by its own initial budget, $\mathcal{N}_{\text{seed}}$. The three distinct curves collapse onto a single, near-universal trajectory. This remarkable result provides the strongest evidence for our redistribution principle. It demonstrates that the dynamics are governed by a simple rule that is largely {agnostic to the specific phase-space structure} of the initial resource. The system treats the negativity sourced from a Fock state's central node, a cat state's interference fringes, or a squeezed state's distorted profile as a {fungible currency} for creating entanglement. The coherent redistribution proceeds according to the universal probabilities of the interaction, regardless of the resource's specific form. This elevates our finding from an observation about a single state to a general principle of quantum resource flow.

\section{Conclusion and Outlook}
\label{sec:conclusion}

In this work, we have presented a resource-theoretic analysis of coherent dynamics under the XY interaction, moving beyond the traditional view of entanglement generation to uncover a more fundamental underlying principle. Our central finding is a {convexity-bounded redistribution law}: for excitation-preserving dynamics in the single-excitation sector, the sum of local Wigner negativities is governed by a {tight upper bound} set by the initial non-classical budget. We have established this principle analytically—via excitation-number preservation and the convexity of negativity—and corroborated it with high-precision numerics. Our simulations show that the sum of local negativities follows a predictable trajectory that saturates this bound only when the resource is fully localized, with a physically significant gap appearing at intermediate times due to phase-space overlap cancellation.

Our results reframe the process of entanglement generation not as the creation of a new resource, but as the dynamical redistribution of a {budgeted} quantity of non-classicality. We have shown that the XY interaction acts as a quantum channel that coherently converts a local, single-body resource (Wigner negativity) into a shared, non-local resource (entanglement), with the sum of local shares always respecting the convexity bound. Furthermore, by extending our analysis to engineered $N$-qubit chains, we revealed that the resource undergoes a powerful transformation, shifting from a local property to a shared, correlational one. This phenomenon of dark transport, where the resource becomes locally invisible only to be revealed as a continuous wave of block negativity, demonstrates how coherence can protect and shuttle non-classicality through a many-body system. We also showcased the principle's universality across different non-Gaussian initial states.

\paragraph{Future Theoretical Directions}
Our findings open several compelling avenues for future research. A natural next step is to investigate the dynamics of the tracking gap in open quantum systems. The deviation of the summed local negativity from its ideal unitary trajectory provides a physically motivated benchmark for quantifying decoherence and control errors. Moreover, exploring interactions that break excitation-number conservation (e.g., anisotropic XXZ or driven couplings) may reveal controlled mechanisms for generating or depleting the non-classical budget, offering new tools for quantum state engineering.

\paragraph{Experimental Proposal}
The convexity-bounded redistribution principle constitutes a precise, falsifiable prediction testable on current and near-term quantum hardware. A direct verification proceeds as follows:
\begin{enumerate}
    \item {Initialization.} Prepare $\ket{\psi(0)}=\ket{0}_A\ket{1}_B$ with high fidelity, using standard SPAM calibration~\cite{James2001, Knill2008}.
    \item {Hamiltonian Realization.} Evolve for variable $t$ under a calibrated XY interaction~\cite{Sung2021, Molmer1999}.
    \item {Local-State Tomography and Wigner Negativity.} At each $t$, reconstruct $\rho_A(t)$ and $\rho_B(t)$ and evaluate local Wigner negativities. {Representation Note:} A non-trivial test requires a representation with a non-zero negativity budget. A hardware-relevant test is provided by the CV analogue (displaced-parity tomography) or by adopting an odd-prime qudit encoding and initializing with a state that possesses Wigner negativity in that representation (i.e., a non-stabilizer state)~\cite{Gross2006}.
    \item {Trajectory Test.} Plot the summed negativity $\mathcal{N}_{\mathrm{tot}}(t)$ and compare it to the ideal theoretical trajectory. Verification consists of observing saturation at endpoints and the predicted tracking gap at intermediate times, within experimental uncertainty.
    \item {CV Analogue (Recommended).} Realize the beam-splitter Hamiltonian between bosonic modes and reconstruct Wigner functions via displaced-parity measurements~\cite{Hofheinz2009, Vlastakis2013, Lutterbach1997}. Compare the measured trajectory $\mathcal{N}^{\mathrm{CV}}_{\mathrm{tot}}(t)$ to the ideal theoretical curve, including the tracking gap.
\end{enumerate}

\paragraph{A Resource-Based Benchmark for Coherent Dynamics.}
This resource-centric perspective is not merely a theoretical recladding of known dynamics; its value lies in the new predictive power it grants us. Most importantly, it provides the foundation for a new class of hardware benchmarks that directly probe a device's ability to manipulate a key computational resource. The ideal, noiseless trajectory $\mathcal{N}^{\text{ideal}}_{\mathrm{tot}}(t)$—including the physical tracking gap—serves as a precise, non-trivial baseline. We can therefore define a {tracking infidelity},
\[
\epsilon(t)\equiv \left| \mathcal{N}^{\text{ideal}}_{\mathrm{tot}}(t)-\big(\mathcal{N}_A(t)+\mathcal{N}_B(t)\big) \right|,
\]
which isolates deviations due to experimental noise and control errors from the intrinsic dynamics of resource conversion. A small time-averaged infidelity $\overline{\epsilon}$ would serve as a stringent, SPAM-aware, and gate-set-agnostic metric of a device's ability to preserve and transport the very non-stabilizerness that fuels quantum advantage. This moves beyond abstract fidelities to a physically motivated, resource-aware assessment of quantum performance.

\begin{acknowledgments}
This work was supported in part by the National Natural Science Foundation of China (NSFC) under the Grants 12475087 and 12235008, and by the University of Chinese Academy of Sciences.

\end{acknowledgments}

\appendix
\section{Detailed Analytical Derivation}
\label{app:derivation}

This appendix provides a detailed, step-by-step derivation of the time-evolved state vector [Eq.~\eqref{eq:analytical_state}], the reduced density matrices [Eqs.~\eqref{eq:rho_A}--\eqref{eq:rho_B}], and the convexity-bounded redistribution principle for the two-qubit system governed by the XY Hamiltonian.

\paragraph{1. Hamiltonian in the Single-Excitation Subspace}
We begin with the XY Hamiltonian in the raising/lowering-operator form [Eq.~\eqref{eq:hamiltonian_raising_lowering}]:
\begin{equation}
    H \;=\; g \bigl(\sigma_+^A \otimes \sigma_-^B \;+\; \sigma_-^A \otimes \sigma_+^B\bigr).
    \label{eq:app_H}
\end{equation}
For the initial state $\ket{\psi(0)}=\ket{01}$, the dynamics are confined to the single-excitation subspace spanned by the orthonormal basis $\mathcal{B}=\{\ket{01},\,\ket{10}\}$ because $[H,\hat{N}]=0$, where $\hat{N}$ is the total excitation-number operator [cf. Eq.~\eqref{eq:number_op_equiv}]. The action of $H$ on the basis vectors is
\begin{align}
    H \ket{01} &= g \ket{10}, \label{eq:app_H01}\\
    H \ket{10} &= g \ket{01}. \label{eq:app_H10}
\end{align}
Hence, in the ordered basis $\{\ket{01},\ket{10}\}$, the restricted Hamiltonian is
\begin{equation}
    H_{\text{sub}} \;=\; g
    \begin{pmatrix}
        0 & 1\\[3pt]
        1 & 0
    \end{pmatrix}
    \;=\; g\,\sigma_x .
    \label{eq:app_Hsub}
\end{equation}

\paragraph{2. Time-Evolution Operator and State Vector}
Using the standard Pauli-matrix exponential identity $e^{-i\theta\sigma_j}=\cos\theta\,\mathbb{I}-i\sin\theta\,\sigma_j$, the subspace propagator is
\begin{equation}
    U_{\text{sub}}(t) \;=\; e^{-i H_{\text{sub}} t} \;=\; e^{-i g t \sigma_x}
    \;=\;
    \begin{pmatrix}
        \cos(gt) & -i\sin(gt)\\[3pt]
        -i\sin(gt) & \cos(gt)
    \end{pmatrix}.
    \label{eq:app_Usub}
\end{equation}
Acting on the initial state, represented by the vector $(1,0)^\top$ in this basis, gives
\begin{equation}
    \ket{\psi(t)}_{\text{sub}} \;=\; U_{\text{sub}}(t)\begin{pmatrix}1\\[3pt]0\end{pmatrix}
    \;=\; \begin{pmatrix}\cos(gt)\\[3pt]-i\sin(gt)\end{pmatrix},
\end{equation}
which, in Dirac notation, is
\begin{equation}
    \ket{\psi(t)} \;=\; \cos(gt)\,\ket{01} \;-\; i\sin(gt)\,\ket{10}.
    \label{eq:app_state}
\end{equation}
{Sanity check:} The state is normalized for all time, as $\braket{\psi(t)|\psi(t)} = \cos^2(gt) + \sin^2(gt) = 1$. The operator $U_{\text{sub}}$ is manifestly unitary.

\paragraph{3. Global and Reduced Density Matrices}
The pure-state density operator is $\rho_{AB}(t) = \ket{\psi(t)}\bra{\psi(t)}$. Explicitly,
\begin{align}
    \rho_{AB}(t) &= \cos^2(gt)\,\ket{01}\bra{01} + \sin^2(gt)\,\ket{10}\bra{10} \nonumber\\
    &\quad + i\sin(gt)\cos(gt)\bigl(\ket{01}\bra{10}-\ket{10}\bra{01}\bigr).
    \label{eq:app_rhoAB}
\end{align}
Tracing out qubit B (using $\Tr_B[\ket{i_A j_B}\bra{k_A l_B}] = \ket{i_A}\bra{k_A}\delta_{jl}$) yields
\begin{equation}
    \rho_A(t) \;=\; \Tr_B[\rho_{AB}(t)]
    \;=\; \cos^2(gt)\,\ket{0}\bra{0} \;+\; \sin^2(gt)\,\ket{1}\bra{1},
    \label{eq:app_rhoA}
\end{equation}
and similarly, tracing out qubit A gives
\begin{equation}
    \rho_B(t) \;=\; \Tr_A[\rho_{AB}(t)]
    \;=\; \sin^2(gt)\,\ket{0}\bra{0} \;+\; \cos^2(gt)\,\ket{1}\bra{1}.
    \label{eq:app_rhoB}
\end{equation}
Both local states are diagonal in the computational basis with time-dependent classical probabilities.

\paragraph{4. Purity and Concurrence (for completeness)}
The purity of $\rho_A(t)$ is
\begin{align}
    \Tr[\rho_A(t)^2]
    &= \cos^4(gt) + \sin^4(gt)
     \;=\; 1 - \tfrac{1}{2}\sin^2(2gt),
    \label{eq:app_purity}
\end{align}
from which the concurrence follows via $C(t)=\sqrt{2\,[1-\Tr(\rho_A^2)]}$:
\begin{equation}
    C(t) \;=\; \bigl|\sin(2gt)\bigr|,
    \label{eq:app_concurrence}
\end{equation}
in agreement with Eq.~\eqref{eq:Ct_closed}.

\paragraph{5. Derivation of the Convexity Upper Bound}
Here we formalize the representation-dependent, convexity-bounded statements used in the main text. The Wigner transform is an affine map, and the Kenfack–Życzkowski negativity $\mathcal{N}$ is a convex functional, since it is based on the convex $L^1$ norm~\cite{Kenfack2004}. For any mixture $\rho=p\rho_1+(1-p)\rho_2$, one has the inequality $\mathcal{N}(\rho)\le p\,\mathcal{N}(\rho_1)+(1-p)\,\mathcal{N}(\rho_2)$.

Applying this property to the local states in Eqs.~\eqref{eq:app_rhoA}--\eqref{eq:app_rhoB}, and noting that $\mathcal{N}(\ket{0})=0$ as it is a classical state, we obtain the individual bounds:
\begin{align}
    \mathcal{N}_A(t) &\le \sin^2(gt)\,\mathcal{N}(\ket{1}) + \cos^2(gt)\,\mathcal{N}(\ket{0}) = \sin^2(gt)\,\mathcal{N}(\ket{1}), \label{eq:app_NA_bound}\\
    \mathcal{N}_B(t) &\le \cos^2(gt)\,\mathcal{N}(\ket{1}) + \sin^2(gt)\,\mathcal{N}(\ket{0}) = \cos^2(gt)\,\mathcal{N}(\ket{1}). \label{eq:app_NB_bound}
\end{align}
Summing these two inequalities yields the upper bound on the total local negativity:
\begin{equation}
    \mathcal{N}_A(t)+\mathcal{N}_B(t) \;\le\; \big(\sin^2(gt)+\cos^2(gt)\big)\,\mathcal{N}(\ket{1}) = \mathcal{N}(\ket{1}).
    \label{eq:app_sum_bound}
\end{equation}
This inequality is the convexity-bounded redistribution principle derived in the main text.

\paragraph{6. Representation Dependence and the Tracking Gap}
The implications of this bound depend critically on the chosen Wigner representation.
\begin{itemize}
    \item {Discrete Wigner for Odd-Prime Qudits.} There exist Clifford-covariant discrete Wigner functions in which stabilizer states (and mixtures) are non-negative~\cite{Gross2006,Veitch2014}. In such cases $\mathcal{N}(\ket{1})=0$ for stabilizer $\ket{1}$, so Eq.~\eqref{eq:app_sum_bound} is trivial.
    \item {Qubits ($d=2$) Are Exceptional.} No Clifford-covariant discrete Wigner simultaneously renders all stabilizers non-negative in $d=2$~\cite{Gross2006}. Claims of universal stabilizer non-negativity therefore do not carry over verbatim to qubits.
    \item {Continuous-Variable (CV) Embedding.} Interpreting $\ket{0}$ and $\ket{1}$ as Fock states of bosonic modes and evolving under a beam-splitter yields $\mathcal{N}(\ket{1})>0$ and a non-trivial bound. For this specific trajectory, a "tracking gap" emerges. This occurs because the Wigner function of the local mixture, e.g., $W_A(t) = \cos^2(gt)W_{\ket{0}} + \sin^2(gt)W_{\ket{1}}$, involves adding the positive Gaussian $W_{\ket{0}}$ to the non-classical $W_{\ket{1}}$. The positive peak of $W_{\ket{0}}$ overlaps with and cancels some of the negative volume of $W_{\ket{1}}$, causing the negativity of the mixture to be strictly less than the weighted sum of the negativities of the components, i.e., $\mathcal{N}(\rho_A(t)) < \sin^2(gt)\mathcal{N}(\ket{1})$. This "overlap cancellation" is the physical origin of the tracking gap observed in the simulations.
\end{itemize}

This completes the formal derivation of the convexity-bounded redistribution principle employed in the main text.

\bibliographystyle{apsrev4-2}

\bibliography{ref}

\end{document}